\documentclass[prl,twocolumn,showpacs,preprintnumbers,amsmath,amssymb]{revtex4}

\usepackage{graphicx}
\usepackage{bm}

\begin{document}

\preprint{Phys. Rev. Lett. (to be published).}

\title{Microwave Penetration Depth and Quasiparticle Conductivity of PrFeAsO$_{1-y}$ Single Crystals: Evidence for a Full-Gap Superconductor}

\author{K. Hashimoto,$^{1}$ T. Shibauchi,$^{1}$ T. Kato,$^{1}$ K. Ikada,$^{1}$ R. Okazaki,$^{1}$ H. Shishido,$^{1}$ \\
M. Ishikado,$^{2}$ H. Kito,$^3$ A. Iyo,$^{3}$ H. Eisaki,$^{3}$ S. Shamoto,$^{2}$ and Y. Matsuda$^{1}$}

\affiliation{$^1$Department of Physics, Kyoto University, Sakyo-ku, Kyoto 606-8502, Japan\\
$^2$Quantum Beam Science Directorate, Japan Atomic Energy Agency, Tokai, Naka, Ibaraki 319-1195, Japan\\
$^3$Nanoelectronics Research Institute (NeRI), National Institute of Advanced Industrial Science and Technology (AIST), 1-1-1 Central 2, Umezono, Tsukuba, Ibaraki 305-8568, Japan}

\date{\today}

\begin{abstract}
In-plane microwave penetration depth $\lambda_{ab}$ and quaiparticle conductivity at 28~GHz are measured in underdoped single crystals of the Fe-based superconductor PrFeAsO$_{1-y}$ ($T_c\approx 35$~K) by using a sensitive superconducting cavity resonator. $\lambda_{ab}(T)$ shows flat dependence at low temperatures, which is incompatible with the presence of nodes in the superconducting gap $\Delta({\bf k})$. The temperature dependence of the superfluid density demonstrates that the gap is non-zero ($\Delta/k_BT_c\gtrsim 1.6$) all over the Fermi surface. The microwave conductivity below $T_c$ exhibits an enhancement larger than the coherence peak, reminiscent of high-$T_c$ cuprate superconductors. 
\end{abstract}


\maketitle


Since the discovery of superconductivity in LaFeAs(O$_{1-x}$F$_x$) \cite{Kam08}, high transition temperatures ($T_c$) up to 56~K have been reported in the doped Fe-based oxypnictides \cite{Tak08,Che08Ce,Ren08Pr,Ren08Nd,Kit08Nd,Che08Sm,Yan08Gd,Wan08Gd}. The nature of superconductivity and the pairing mechanism in this system are fundamental physical problem of crucial importance. The first experimental task to this problem is to elucidate the superconducting pairing symmetry, which is intimately related to the pairing interaction.

The NMR Knight-shift measurements appear to indicate the spin-singlet pairing \cite{Gra08,Mat08}. However, the superconducting gap structure, particulary the presence or absence of nodes in the gap, is highly controversial. The specific heat shows a nonlinear magnetic field dependence \cite{Mu08}.  The NMR relaxation rate shows the absence of the coherence peak and the $T^3$-dependence below $T_c$ \cite{Nak08,Mat08,Gra08,Muk08}.  The lower critical field exhibits a $T$-linear dependence at low temperatures \cite{Ren08Hc1}.  The $\mu$SR experiments report an unusual field-dependence of the penetration depth \cite{Lue08}.  In the point-contact spectroscopy, a zero-bias conductance peak is reported \cite{Sha08,Wan08}. These results have been interpretated as an indication of unconvensional superconductivity with line nodes.  On the other hand, the Andreev reflection data are found to be consistent with an isotropic gap \cite{Che08}.   All of these experiments have been performed by using polycrystalline samples.   Definitely, measurements using single crystals are highly desired to obtain unambiguous conclusions on the superconducting gap structure.

In this paper, we report on the measurements of the complex surface impedance in underdoped single crystals of the oxypnictide superconductor PrFeAsO$_{1-y}$ ($T_c\approx35$~K), from which properties of thermally excited quasiparticles can be directly deduced. Since the recent NMR experiments of the Pr-based iron oxypnictide suggest the non-magnetic state in the superconducting samples \cite{Mat08}, PrFeAsO$_{1-y}$ seems suitable for the penetration depth study \cite{Coo96,Pro06}.  Moreover, PrFeAsO$_{1-y}$ has a higher $T_c$ than that of La-compounds, which enables the measurements in a wider temperature range.  We observe flat temperature dependence of the in-plane penetration depth $\lambda_{ab}(T)$ at low temperatures, indicating exponentially small quasiparticle excitations, which clearly contradicts the presence of nodes in the gap.  The quasiparticle conductivity is enhanced compared with the BCS theory below $T_c$, which bears a similarity with high-$T_c$ cuprates and heavy fermion superconductors with strong electron scattering in the normal state above $T_c$.

At microwave frequencies $\omega$, the surface impedance $Z_s=R_s+{\rm i}X_s$, where $R_s$ ($X_s$) is the surface resistance (reactance), and complex conductivity $\sigma=\sigma_1-{\rm i}\sigma_2$ are given by 
\begin{equation}
Z_s=R_s+{\rm i}X_s=\left(\frac{{\rm i}\mu_0\omega}{\sigma_1-{\rm i}\sigma_2}\right)^{1/2}
\label{impedance}
\end{equation}
for the case of local electrodynamics.  In the superconducting state, the surface reactance provides a direct measure of the London penetration depth $\lambda$ via $X_s(T)=\mu_0\omega\lambda(T)$.  The number of excited quasiparticles is most directly related to $\lambda(T)$, since the superfluid density $n_s$ is proportional to $\lambda^{-2}$. The real part of conductivity $\sigma_1$ is determined by the quasiparticle dynamics, and in the simple two-fluid model, which is known to be useful in superconductors with strong correlations \cite{Bon93}, $\sigma_1$ is related to the quasiparticle scattering time $\tau$ through $\sigma_1=(n-n_s)e^2\tau/m^*(1+\omega^2\tau^2)$, where $n$ is the total density of carriers with effective mass $m^*$.


Single crystals of PrFeAsO$_{1-y}$ were grown by high-pressure synthesis method using a belt-type anvil apparatus. Powders of PrAs, Fe, Fe$_2$O$_3$ were used as the starting materials. PrAs was obtained by reacting Pr chips and As pieces at 500$^\circ$C for 10 hours and then 850$^\circ$C for 5 hours in an evacuate quarts tube. The starting materials were mixed at nominal compositions of PrFeAsO$_{0.6}$, and pressed into pellets. The samples were grown by heating the mixtures in BN crucibles under a pressure of about 2~GPa at 1300$^\circ$C for 2 hours. Plate-like crystals up to $150\times150~\mu$m$^2$ in the $ab$ planes are extracted mechanically and they have shiny surfaces. Our crystals, whose $T_c$ ($\approx 35$~K) is lower than the optimum $T_c\approx51$~K in PrFeAsO$_{1-y}$ \cite{Ren08}, are in the underdoped regime, which is close to the spin density wave order \cite{Zha08}. We have carefully checked the homogeneity of the crystals by using the magneto-optics measurements of field distribution \cite{Oka08}, which reveal that a nearly perfect homogeneous Meissner state is attained within 2~K below $T_c$. 

\begin{figure}[t]
\includegraphics[width=92mm]{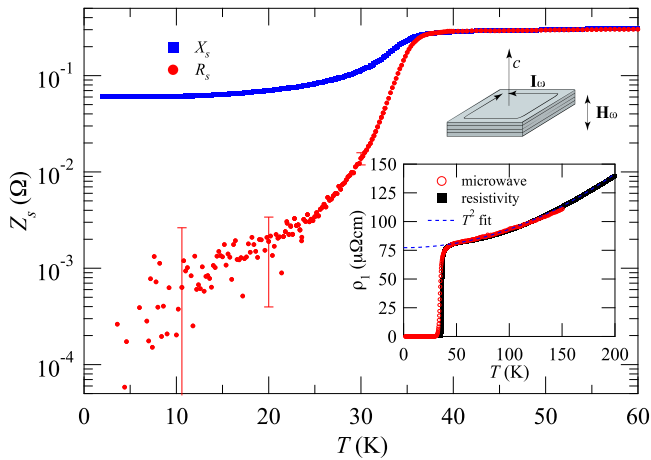}%
\caption{(color online). Temperature dependence of the surface resistance $R_s$ and reactance $X_s$ at 28~GHz in a PrFeAsO$_{1-y}$ single crystal (\#1). In the normal state, $R_s=X_s$ as expected from Eq.~(\ref{impedance}). The low-temperature errors in $R_s$ are estimated from run-to-run uncertainties in $Q$ of the cavity. Inset shows the  microwave resistivity $\rho_1(T)$ (red circles) compared with the dc resistivity in a crystal from the same batch (black squares). The dashed line represents a $T^2$ dependence. The measurement configuration is also sketched. 
}
\label{Z_s}
\end{figure}

The surface impedance was measured by a cavity perturbation method with the hot finger technique \cite{Shi94,Shi07}. We used a 28~GHz TE$_{011}$-mode superconducting Pb cavity with a high quality factor $Q\sim10^6$. To measure the surface impedance of the small single crystal with high precision, the cavity resonator is soaked in the superfluid $^4$He at 1.6~K and its temperature is stabilized within $\pm1$~mK.   We place a crystal in the antinode of the microwave magnetic field ${\bf H}_\omega$ ($\parallel c$ axis) so that the shielding current ${\bf I}_\omega$ is excited in the $ab$ planes [see sketch in Fig.~\ref{Z_s}]. The inverse of quality factor $1/Q$ and the shift in the resonance frequency are proportional to $R_s$ and the change in $X_s$, respectively. To determine the absolute values of $Z_s$, the in-plane dc resistivity $\rho(T)$ was measured by using the W deposition technology of a focused ion beam system to put 4 contacts. We also use $\lambda_{ab}(0)= 280$~nm, which is determined from the lower critical field measurements using a micro-array of Hall probes in the crystals from the same batch \cite{Oka08}. In our frequency range, the skin depth $\delta_{cl}$ is much shorter than the sample width, ensuring the so-called skin-depth regime. 


\begin{figure}[t]
\includegraphics[width=92mm]{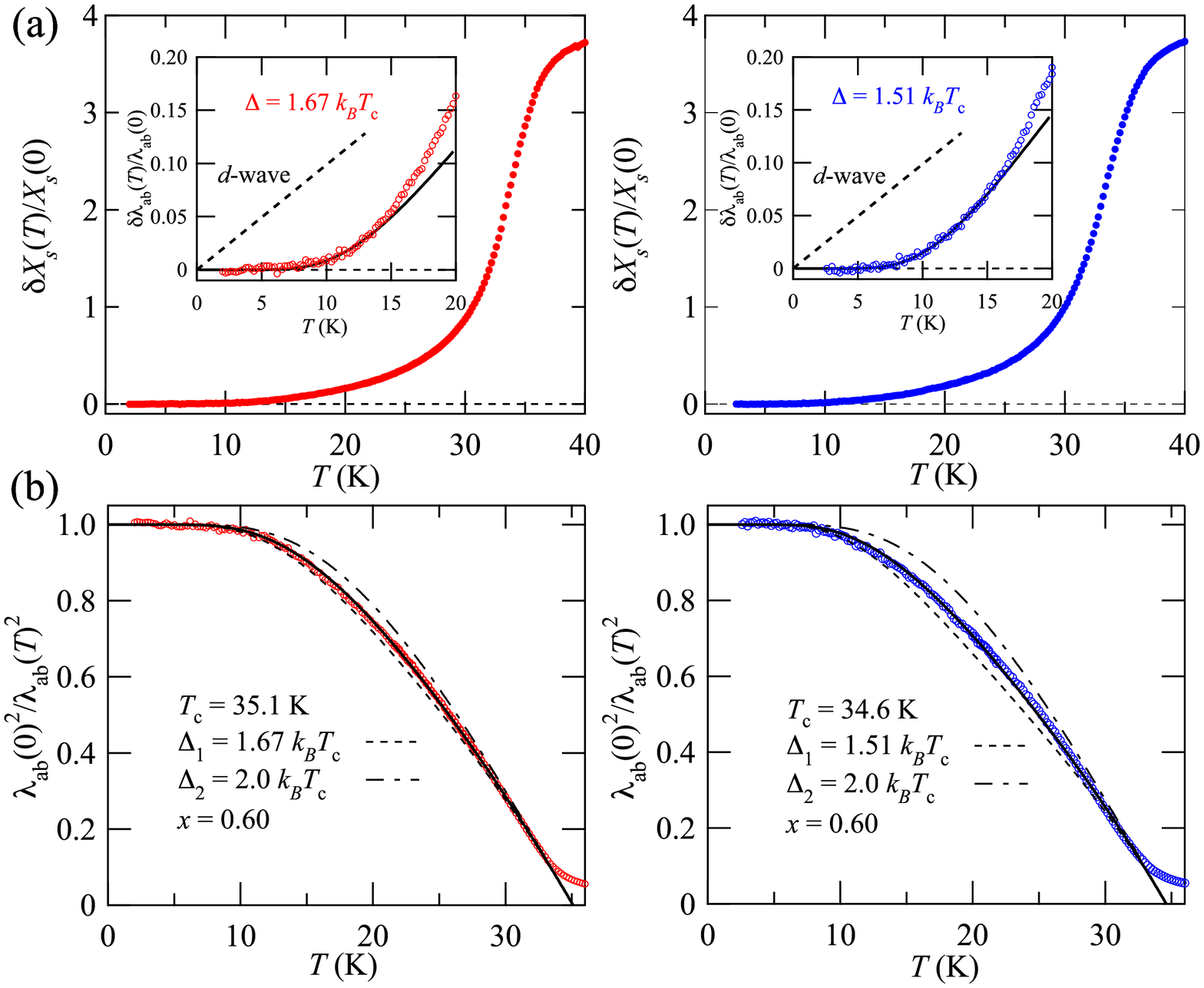}%
\caption{(color online). (a) Temperature dependence of $\delta X_s(T)/X_s(0)$, which corresponds to $\delta\lambda_{ab}(T)/\delta\lambda_{ab}(0)$ in the superconducting state, for crystals \#1 (left) and \#2 (right). Note that the experimental $X_s$ is limited by $\mu_0\omega\delta_{cl}/2$ above $T_c$. Inset is an expanded view at low temperatures. The dashed lines represent $T$-linear dependence expected in clean $d$-wave superconductors with line nodes. The solid lines are low-$T$ fits to Eq.~(\ref{BCS}). (b) Temperature dependence of the superfluid density $\lambda_{ab}^2(0)/\lambda_{ab}^2(T)$. The solid lines are the best fit results to the two-gap model, and the dashed and dashed-dotted lines are the single gap results for $\Delta_1$ and $\Delta_2$, respectively. $T_c$ is defined by the temperature at which the superfluid density becomes zero.} 
\label{lambda}
\end{figure}

Figure~\ref{Z_s} shows typical temperature dependence of the surface resistance $R_s$ and $X_s$. In the normal state where $\omega\tau$ is much smaller than unity, the temperature dependence of microwave resistivity $\rho_1={2R_s^2/\mu_0\omega}$ is expected to follow $\rho(T)$ [see Eq.~(\ref{impedance})]. Such a behavior is indeed observed in the inset of Fig.~\ref{Z_s}. Below about 100~K $\rho_1(T)$ exhibits a $T^2$ dependence and it shows a sharp transition at $T_c\approx 35$~K. As shown in the main panel of Fig.~\ref{Z_s}, the crystal has low residual $R_s$ values in the low temperature limit. These results indicate high quality of the crystals. We note that the transition in microwave $\rho_1(T)$ is intrinsically broader than that in dc $\rho(T)$, since the applied 28-GHz microwave (whose energy corresponds to 1.3~K) excites additional quasiparticles just below $T_c$. 

In Fig.~\ref{lambda}(a) shown are the normalized change in the surface reactance $\delta X_s(T)=X_s(T)-X_s(0)$, and the corresponding normalized change in the in-plane penetration depth $\delta\lambda_{ab}(T)=\lambda_{ab}(T)-\lambda_{ab}(0)$ for two crystals. The overall temperature dependence of $\delta X_s(T)$ in these crystals is essentially identical, which indicates good reproducibility. It is clear from the inset of Fig.~\ref{lambda}(a) that $\delta\lambda_{ab}(T)$ has flat dependence at low temperatures. First we compare our data with the expectations in unconventional superconductors with nodes in the gap. In clean superconductors with line nodes, thermally excited quasiparticles near the gap nodes give rise to the $T$-linear temperature dependence of $\delta\lambda_{ab}(T)$ at low temperatures, as observed in YBa$_2$Cu$_3$O$_{7-\delta}$ crystals with $d$-wave symmetry \cite{Har93}. In the $d$-wave case, $\delta\lambda_{ab}(T)/\lambda_{ab}(0)\approx \frac{\ln2}{\Delta_0}k_BT$ is expected \cite{Pro06}, where $\Delta_0$ is the maximum of the energy gap $\Delta({\bf k})$. This linear temperature dependence with an estimation $2\Delta_0/k_BT_c\approx4$ \cite{Nak08} [dashed line in Fig.~\ref{lambda}(a)] distinctly deviates from our data. When the impurity scattering rate $\Gamma_{\rm imp}$ becomes important in superconductors with line nodes, the induced residual density of states changes the $T$-linear dependence into $T^2$ below a crossover temperature $T^*_{\rm imp}$ determined by $\Gamma_{\rm imp}$ \cite{Pro06}. This is also clearly different from our data in Fig.~\ref{lambda}(a). If by any chance the $T^2$ dependence with a very small slope should not be visible by the experimental errors below $\sim10$~K, then we would require enormously high $T^*$. However, since no large residual density of states is inferred from NMR measurements even in polycrystalline samples of La-system with lower $T_c$ \cite{Nak08,Muk08}, such a possibility is highly unlikely. These results lead us to conclude that in contradiction to the presence of nodes in the gap, the finite superconducting gap opens up all over the Fermi surface.

In fully gapped superconductors, the quasiparticle excitation is of an activated type, which gives the exponential dependence 
\begin{equation}
\frac{\delta\lambda_{ab}(T)}{\lambda_{ab}(0)} \approx \sqrt{\frac{\pi\Delta}{2k_{\rm B}T}}\exp\left(-\frac{\Delta}{k_{\rm B}T}\right)
\label{BCS}
\end{equation}
at $T\lesssim T_c/2$ \cite{Hal71}. Comparisons between this dependence and the low-temperature data in Fig.~\ref{lambda}(a) enable us to estimate the minimum energy $\Delta_{\rm min}$ required for quasiparticle excitations at $T=0$~K; {\it i.e.} $\Delta_{\rm min}/k_BT_c=1.6\pm0.1$. 

We can also plot the superfluid density $n_s= \lambda_{ab}^2(0)/ \lambda_{ab}^2(T)$ as a function of temperature in Fig.~\ref{lambda}(b). Again, the low-temperature behavior is quite flat, indicating a full-gap superconducting state. We note that by using the gap $\Delta_{\rm min}$ obtained above alone, we are unable to reproduced satisfactory the whole temperature dependence of $n_s$ [see the dashed lines in Fig.~\ref{lambda}(b)], although a better fit may be obtained by using a larger value of $\Delta=1.76k_BT_c$, the BCS value. Since Fe oxypnictides have the multi-band electronic structure \cite{Sin08,Liu08}, we also try to fit the whole temperature dependence with a simple two-gap model $n_s(T)=xn_{s1}(T)+(1-x)n_{s2}(T)$: The band 1 (2) has the superfluid density $n_{s1}$ ($n_{s2}$) which is determined by the gap $\Delta_1$ ($\Delta_2$), and $x$ defines the relative weight of each band to $n_{s}$. This simple model was successfully used for the two-gap $s$-wave superconductor MgB$_2$ with a large gap ratio $\Delta_2/\Delta_1\approx2.6$ \cite{Fle05}. Here we fix $\Delta_1=\Delta_{\rm min}$ obtained from the low-temperature fit in Fig.~\ref{lambda}(a), and excellent results of the fit are obtained for $\Delta_{2}/k_BT_c=2.0$ and $x=0.6$ [see Fig.~\ref{lambda}(b)]. These exercises suggest that the difference in the gap value of each band in this system is less substantial than the case of MgB$_2$. 

\begin{figure}[t]
\includegraphics[width=80mm]{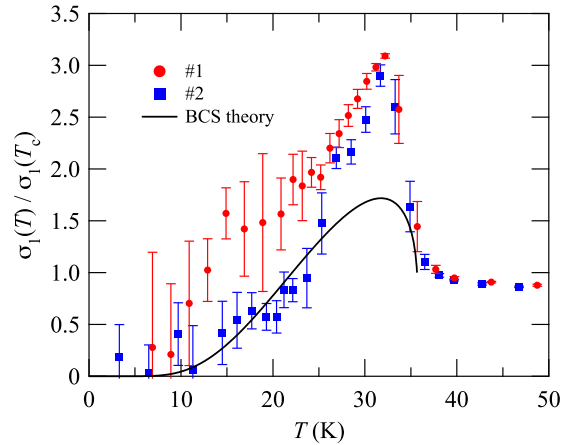}%
\caption{(color online). Temperature dependence of quasiparticle conductivity $\sigma_1$ normalized by its $T_c$ value at 28~GHz for two crystals. The solid line is a BCS calculation \cite{Zim91} of $\sigma_1(T)/\sigma_1(T_c)$ with $\tau=1.2\times10^{-13}$~s, which is estimated from $\rho(T_c)=77~\mu\Omega$cm and $\lambda_{ab}(0)=280$~nm.
} \label{sigma}
\end{figure}

Next let us discuss the quasiparticle conductivity $\sigma_1(T)$, which is extracted from $Z_s(T)$ through Eq.~(\ref{impedance}). The results for samples \#1 and \#2 are demonstrated in Fig.~\ref{sigma}. Although at low temperatures we have appreciable errors, it is unmistakable that $\sigma_1(T)$ shows a large enhancement just below $T_c$ \cite{fluctuation}. This enhancement is considerably larger than the coherence peak expected in the BCS theory \cite{Zim91}. Similar large and broad enhancements of $\sigma_1(T)$ in the superconducting state have been observed in the high-$T_c$ cuprate \cite{Bon93} and heavy-Fermion superconductors \cite{Orm02} having strong inelastic electron scattering that is gapped below $T_c$. There, the enhancement has been attributed to the strong suppression of the quasiparticle scattering rate $1/\tau$. It warrants further studies to clarify whether the observed enhancement of $\sigma_1(T)$ in PrFeAsO$_{1-y}$ comes from the same origin, or from some novel mechanism that enhances the coherence peak in the multiband system.

Various theories have been proposed for the pairing symmetry in the doped Fe-based oxypnictides \cite{Maz08,Kur08,Lee08,Si08,Tes08,Sta08,Seo08,Wan08Lee}. Among them, theories based on antiferromagnetic spin fluctuations \cite{Maz08,Kur08} predict fully gapped superconductivity although there is a sign reversal of the order parameter between different Fermi surface sheets. As confirmed by recent ARPES measurements \cite{Liu08}, doped Fe-based oxypnictides have hole pockets in the Brillouin zone center and electron pockets in the zone edges \cite{Sin08}. It has been suggested that the nesting vector between these pockets is important, which favors an extended $s$-wave order parameter having opposite signs between the hole and electron pockets. Our penetration depth result of full gap is in good correspondence with such an extended $s$-wave state (or $s_{\pm}$ state) with no nodes in both gaps in these two bands. In addition, such a sign change would suppress the coherence peak in the NMR relaxation rate, in agreement with the experiments \cite{Nak08,Mat08,Gra08,Muk08}. We note that other theories \cite{Tes08,Sta08,Seo08} are also consistent with the fully gapped superconductivity. 

In summary, we provide strong evidence for the absence of nodes in the superconducting gap from microwave penetration depth measurements in non-magnetic PrFeAsO$_{1-y}$ single crystal with a sharp transition at $T_c\approx35$~K. The flat dependence of $\lambda_{ab}(T)$ and $n_s(T)$ at low temperatures demonstrates that the finite superconducting gap larger than $\sim1.6k_BT_c$ opens up all over the Fermi surface. The microwave conductivity exhibits an enhancement larger than the BCS coherence peak, reminiscent of superconductors with strong electron scattering. The present results impose an important constraint on the order parameter symmetry, namely the newly discovered Fe-based high-$T_c$ superconductors are fully gapped in contrast to the high-$T_c$ cuprate superconductors. 


We thank S. Fujimoto, H. Ikeda, K. Ishida, and H. Kontani for fruitful discussion, and M. Azuma, S. Kasahara, K. Kodama, T. Saito, Y. Shimakawa, T. Terashima, M. Yamashita, and K. Yoshimura for technical assistance. This work was supported by KAKENHI (No. 20224008) from JSPS, by Grant-in-Aid for the global COE program ``The Next Generation of Physics, Spun from Universality and Emergence'' and by Grant-in-Aid for Specially Promoted Research (No. 17001001) from MEXT, Japan. 

After submission, MHz penetration depth measurements have been reported in SmFeAsO$_{1-x}$F$_y$ \cite{Mal08} and NdFeAsO$_{0.9}$F$_{0.1}$ \cite{Mar08}, which are consistent with our conclusion of a full-gap superconducting state in Fe-oxypnictides.

\end{document}